# Band alignment of two-dimensional transition metal dichalcogenides: application in tunnel field effect transistors

Cheng Gong,[1] Hengji Zhang,[2] Weihua Wang,[1] Luigi Colombo,[3] Robert M. Wallace,[1,2] and Kyeongjae Cho[1,2,*]

[1]Department of Materials Science and Engineering, The University of Texas at Dallas, Richardson, TX 75080, USA

[2]Department of Physics, The University of Texas at Dallas, Richardson, TX 75080, USA

[3]Texas Instruments, Dallas, Texas 75243, USA

Tunnel field effect transistors (TFETs) based on vertical stacking of two dimensional materials are of interest for low-power logic devices. The monolayer transition metal dichalcogenides (TMDs) with sizable band gaps show promise in building p-n junctions (couples) for TFET applications. Band alignment information is essential for realizing broken gap junctions with excellent electron tunneling efficiencies. Promising couples composed of monolayer TMDs are suggested to be Ⅵ B-MeX$_2$ ( Me= W, Mo; X= Te, Se) as the *n*-type source and  ⅠVB-MeX$_2$ (Me = Zr, Hf; X= S, Se) as the *p*-type drain by density functional theory calculations.

[*]Author to whom correspondence should be addressed. Electronic mail: kjcho@utdallas.edu

As the scaling of conventional metal-oxide-semiconductor field effect transistors (MOSFETs), a number of challenges emerge including degraded carrier mobility, short channel effects, and interfacial defect states which affect the operational speed, device stability, and the abrupt switching negatively. The design and development of alternative electronic devices is urgently needed and has been persistently pursued in recent years. The TFET is a promising device candidate for future low-energy electronic systems, of which two characteristic parameters needs careful consideration: on-off current ratio and subthreshold swing (SS).[1] Tunnel FETs based on III-V materials have been explored with promising results.[2] However, a major challenge for adopting these materials is the control of interfacial defects, such as dangling bonds, and the resultant interface state density which impacts the SS.[3, 4]

In contrast, 2D materials, with their inherent surface bonding properties, provide an opportunity to substantially reduce interface state densities, and thus mitigate SS degradation. Tunnel FETs based on vertical stacking of graphene with atomically thin hexagonal boron nitride ($h$-BN) or molybdenum disulfide ($MoS_2$) layers as separation barriers have been experimentally demonstrated with precise sub-nanoscale structures.[5] However, even though the switching function is realized, the zero gap nature of graphene is still a limiting factor for increasing the on-off ratios.

Band-to-band tunneling transport can be utilized to explore heterogeneous TFETs composed of stacked source-drain junctions.[6] Semiconducting 2D materials are of interest facilitating high electron tunneling efficiency through a "broken-gap" band alignment while suppressing the OFF current by band gap engineering of drain materials. Owing to the availability of a range of electronic band gaps, many 2D TMDs can be considered as promising constituent materials for TFET applications, since (1) the size of the band gap is tunable, *e.g.*, by mechanical strain[7] or the composition control in ternary compounds, (2) the Fermi level can be effectively adjusted by doping,[8] and (3) both band gap and Fermi level can be simultaneously adjusted by layer stacking.[9]

This work focuses on the systematic density functional theory (DFT) calculation of individual isolated 2D TMDs. Considering the large family of TMDs, with an even larger number of possible TMD stacked *p-n* junctions ("couples"), *ab initio* calculations of all TMD couples is a burdensome task. Moreover, the non-trivial lattice mismatches among most TMDs provide a formidable challenge for DFT. Therefore, the study of individual isolated 2D TMDs is the first important step for finding clues on selecting couples suitable for TFETs. Such information will also aid in providing focus to the growth of high quality 2D TMD materials, essential for accurate device studies. In particular, the information about the universal band alignment of TMDs with respect to the vacuum level is of practical significance.

We examine, by DFT calculations, the electronic properties of 24 types of 2D TMDs consisting of the combinations between IVB-VIB metals (IVB: Ti, Zr and Hf; VB: V, Nb, and Ta; and VIB: Mo and W) and chalcogen species (S, Se, and Te), and find that their electronic properties vary dramatically. Semiconductors are found in all VIB- and some IVB-TMDs, with VIB-TMDs work functions in the range of 4 to 5 eV and IVB-TMDs work functions at about 6 eV. The large work function difference between VIB- and IVB-TMDs provides a way of forming steep junctions with broken-gap band alignment. All of the calculated 2D VB-TMDs display metallic character which will not be discussed in this Letter.

The calculations are performed by VASP[10] with projector-augmented wave (PAW)[11] pseudopotential with both generalized gradient approximation (GGA) of Perdew-Burke-Ernzerhof (PBE)[12] functional and local density approximation (LDA)[13] to describe the exchange-correlation. Spin polarization is applied for both ionic and electronic relaxation, and spin-orbit coupling (SOC) is switched on for band structure calculations. In the first step, all the studied bulk TMDs are relaxed based on both 2$H$- and 1$T$-crystal structures, with the remnant force on each atom below 0.01 eV/Å as the stopping criterion. Based on the optimized bulk lattice constants, monolayer structures are set with a vacuum region of 17 Å normal to the surface and further relaxed with the fixed unit cell shape and size. The Monkhorst-Pack k-point sampling in Brillouin zone (BZ) is Γ-centered with 8×8×1 and 40×40×1 meshes in ionic and electronic optimization, respectively. The energy cutoff is chosen at 400 eV, and the electronic optimization stops when the total energies of neighboring optimization loops differ below $10^{-4}$ eV.

Reliable band alignment information of semiconductors is based on the accurate prediction of two quantities: the size of band gap and the absolute energy of the Fermi level. There is a recent debate about the electronic band gap of monolayer $MoS_2$. The value of 1.88 eV obtained by photoluminescence (PL) measurements[14] is in question and thought to be an exitonic gap,[15, 16] rather than a transport gap. Recent calculations incorporating the GW approximation[17] collectively claim a much larger quasiparticle gap of about 2.8 eV with a strong exciton binding energy of about 1.0 eV, resulting in an excitonic gap of about 1.8 eV. Experimental verification of the 2.8 eV transport gap has not yet been demonstrated. In addition, the calculated electron affinity of monolayer $MoS_2$ by PBE is 4.27 eV, in this work and others.[18, 19] However, the experimentally reported electron affinity in bulk $MoS_2$ crystals is about 4.0 eV,[20] and it is known that an enlarged band gap with decreased electron affinity is expected to be found in $MoS_2$ of decreased number of layers.[18, 21] Consequently, employing a GW correction in the calculation is necessary for gaining absolute energies of band edges.

We carry out "single-shot $G_0W_0$" on the optimized MoS$_2$ structures by PBE with 12×12×1 k-point sampling in the irreducible BZ, and a 3.0 eV quasiparticle energy gap is obtained, verifying the substantially larger band gap. Converged calculation of the absolute quasiparticle band edge energies requires at least ~1000 conduction bands included in $G_0W_0$ calculation, reported recently by Liang *et al.*.[19] The GW correction of the conduction band minimum (CBM) and valence band maximum (VBM) was claimed to obey the band-gap-center approximation, which means that there is about the same amount of shift in inverse directions of both band edges relative to the same band-gap-center. According to the report in Ref. 19, the gap opening by the GW correction is ~50% for monolayers of Mo and W TMDs. Here, we do not extend the costly GW calculations to all the studied systems, but include a GW correction of 50% gap opening with the band gap center (*i.e.*, Fermi level) unchanged. The minor difference between gap sizes and edge energies of our results and Liang *et al.*'s results is mainly because spin-orbit coupling (SOC) is included in our study. Based on the universal band alignments of all the studied semiconducting 2D TMDs, promising material couples for TFETs are suggested.

The total energy calculations of the optimized monolayer TMDs based on both PBE and LDA show that all VIB-TMDs favor a trigonal prismatic coordination, whereas IVB-TMDs favor an octahedral coordination as summarized in Table I. For convenience, the trigonal prismatic coordination in monolayer is denoted as "*H*" (as opposed to the bulk terminology "2*H*"), and the octahedral coordination in monolayer as "*T*" (as opposed to the bulk terminology "1*T*"). The preferred coordination can be understood based on the electronic configuration of TM atoms and

110    the crystal field splitting. For VIB-TM atoms, the six outermost valence electrons can bond to six

111    chalcogen atoms through the trigonal prismatic coordination. As shown in Fig. 1, the effect on the

112    metal *d*-levels of a trigonal prismatic ligand field is splitting off of the $d_z^2$ subband at lower

113    energy.[22] VIB-TM atoms have two electrons for filling the subband and stabilize the trigonal

114    prismatic coordination, but IVB-TM atoms do not have the two electrons to fill the $d_z^2$ subband so

115    that an octahedral coordination is preferred.

116        The band gaps calculated by PBE-SOC and corrected by GW based on the band-gap-center

117    approximation are plotted in Fig. 2. For the *H*-monolayer VIB-TMDs, as the atomic indices of

118    chalcogen species increase from S to Te, the valence band edge undergoes a conspicuous energy

119    increase, associated with a relatively smaller energy increase of conduction band edge, resulting in

120    a decreasing energy gap. As the atomic indices of chalcogen atoms increase, the larger atomic

121    radius and decreased reactivity induce weakened inter-atomic interaction strength and a larger

122    lattice constant. Thus, a smaller band gap is produced. For the same chalcogen species, Mo is

123    more reactive than W because of the intrinsic higher reactivity of 3*d*-electrons than 4*d*-electrons.

124    Hence, the overall energy levels of Mo-dichalcogenides are lower than that of W-dichalcogenides.

125        In contrast, IVB-TM atoms have one less pair of valence electrons. As a result, IVB-TMDs

126    are also semiconducting with a deeper band with respect to VIB-TMDs. Hence, the work

127    functions of IVB-TMDs are larger compared with those of VIB-TMDs. The intralayer bonding in

128    IVB-TMDs is more ionic and weaker, exhibiting generally smaller band gaps. The size of the gap

129    is strongly dependent on the TM and chalcogen species. The smaller band gaps are associated with

130    higher atomic indices of constituent chalcogen atoms, producing negative gaps in $HfTe_2$ and $ZrTe_2$,

131    which are predicted to be unsuitable for TFET applications. This trend is due to the decreasing electron negativity of the higher-indexed chalcogen species. Semiconducting IVB-TMDs with non-trivial band gaps are found in $ZrS_2$, $ZrSe_2$, $HfS_2$ and $HfSe_2$.

The presence of finite band gaps but distinctive positions of band edges of VIB- and IVB-TMDs imply the potential combination of these two groups of TMDs for TFET applications. As indicated by results of Fig. 2, electrons at the valence band edges of $WTe_2$ and $MoTe_2$ would be able to tunnel into the conduction bands of $ZrS_2$, $ZrSe_2$, $HfS_2$ and $HfSe_2$ with ease. While replacing Te species by Se in VIB-TMDs, the efficiency for electron injection from VIB-TMDs to IVB-TMDs decreases due to the lower valence band edges of VIB-selenides. The VBMs of $MoS_2$ and $WS_2$ are even lower than the CBMs of IVB-TMDs, which is detrimental for "broken-gap" alignment formation and not promising for TFET applications. Considering the relative energy level shift[23] for TMDs in direct contact or in contact with dielectric media, the final band alignment in such systems needs further examination. The calculation herein provides useful guidance for selecting the TMD material couples for TFETs.

It is interesting to note the intrinsic scattering in the suggested TFETs integrating VIB-TMDs as the *n*-type source and IVB-TMDs as the *p*-type drain. As shown in Fig. 3, *H*-monolayer VIB-TMDs have direct band gaps at K point, whereas indirect gaps are formed between $\Gamma$ and M points for *T*-monolayer IVB-TMDs. Correspondingly, the electron tunneling from the VBM of the *H*-monolayer VIB-TMDs at the K-point to the CBM of the *T*-monolayer IVB-TMDs at the M-point is anticipated to experience inelastic scattering. Engineering the position of band edges in reciprocal space is a feasible scheme for addressing the issue of intervalley scattering. To this end, we analyze the band edge properties of monolayer VIB- and IVB-TMDs. Insights on how strain affects the band properties of monolayer TMDs are obtained.

For *H*-monolayer VIB-TMDs, we choose to study $MoS_2$ as a representative for a convenient comparison with other published data. As shown in Fig. 3(a), the bonding states at $\Gamma(V)$ and $K(V)$ are mainly composed of out-of-plane $d_z^2$-orbitals and in-plane $d(x^2-y^2)+d_{xy}$ orbitals, respectively. Therefore, the in-plane tensile strain weakens the inter-atomic $d_z^2$-$d_z^2$ bonding more than the $d(x^2-y^2)$-$d(x^2-y^2)$ and $d_{xy}$-$d_{xy}$ bondings. The resulting energy level shifting up of the $d_z^2$-$d_z^2$ bonding state $\Gamma(V)$ is more than that of the bonding state $K(V)$. For the anti-bonding states, $K(C)$ is composed of a higher weight of out-of-plane oriented orbitals than $\Gamma K(C)$. Hence, the in-plane tensile strain shifts the energy position of $K(C)$ downward in a larger amount compared with that of $\Gamma K(C)$. The tensile strain is inferred to produce a reduced indirect band gap between $\Gamma(V)$ and $K(C)$, whereas the compressive strain is expected to produce an increased indirect band gap between $K(V)$ and $\Gamma K(C)$. These findings agree well with the reported strain-induced band gap evolution by first-principle simulations.[7, 9, 24, 25]

For *T*-monolayer IVB-TMDs, $ZrS_2$ is studied as a representative system. The intrinsic band gap is indirect between $\Gamma$ as VBM and M as CBM. Interestingly, as shown in Fig. 3(b), among $\Gamma(C)$, $\Gamma(V)$, $M(C)$ and $M(V)$, only $M(C)$ has a significant portion of our-of-plane oriented orbitals Zr-$dz^2$ and S-$pz$. Hence, strain is expected to affect the position of $M(C)$ more than that of others. We apply 5% tensile and compressive strains, and find that $M(C)$ shifts up with tensile strain and shifts down with compressive strain, but $\Gamma(C)$, $\Gamma(V)$ and $M(V)$ do not change much. Upon the application of larger than 5% tensile strain, the CBM begins to transit from M to $\Gamma$. Tensile strains are indicated to be able to modify both the VBM of 2D VIB-TMDs and the CBM of 2D IVB-TMDs to $\Gamma$ point, potentially eliminating the inelastic scattering in tunneling process.

In summary, by systematic DFT calculations, we found semiconductors in 2D form of both VIB- and IVB-TMDs. The two groups of semiconductors have distinct work functions, exhibiting the potential of being stacked for TFET applications. With more accurate prediction of the quasiparticle energy gap by GW corrections, promising 2D TMDs couples still exist. We suggest the layer coupling using VIB-MeX$_2$ (Me= W and Mo; X= Te and Se) as the *n*-type source and IVB-MeX$_2$ (Me = Zr and Hf; X= S and Se) as the *p*-type drain, for vertically stacked TFET applications. The two groups of semiconductors have distinct band edge characters, which intrinsically incur intervalley scattering during the electron tunneling process. Strain is highlighted as an effective way to modify the band edge properties of these 2D TMDs. Interface related issues need to be carefully examined in realistic TFET designs and fabrication where the work functions of 2D materials is inevitably modified by contact to surrounding media. In addition, the conclusion about the preferential couples in this Letter is based on the stacking of monolayers. Heterogeneous junctions between multilayer TMDs require further study, and the precise band realignment caused by interface charge transfer and interface dipole formation when two isolated monolayers couple together, will be the focus of a follow-up work. At the current stage of research, the fundamental understanding of the TMD family provided here will inspire exciting ideas and applications.

This work was supported by the Center for Low Energy Systems Technology (LEAST), one of six centers supported by the STARnet phase of the Focus Center Research Program (FCRP), a Semiconductor Research Corporation program sponsored by MARCO and DARPA. HZ is partially supported by Nano·Material Technology Development Program through the National Research Foundation of Korea (NRF) funded by the Ministry of Science, ICT and Future Planning (2012M3A7B4049888).

TABLE I. Optimized lattice constants and total energies of monolayer TMDs, calculated by both PBE and LDA (in parenthesis). $a_H$ and $a_T$ are the in-plane lattice constant of trigonal prismatic and octahedral monolayer TMDs respectively. ΔE is the energy difference between trigonal prismatic and octahedral structures. Negative ΔE means the *H*-structure is more stable with lower total energy, and vice versa.

|    | S | Se | Te |
|----|---|----|----|
| Mo | $a_H$= 3.19 (3.12) Å | $a_H$= 3.33 (3.25) Å | $a_H$= 3.56 (3.47) Å |
|    | ΔE=-0.838 (-0.864) eV | ΔE=-0.706 (-0.709) eV | ΔE=-0.515 (-0.504) eV |
| W  | $a_H$= 3.19 (3.13) Å | $a_H$= 3.32 (3.25) Å | $a_H$= 3.56 (3.48) Å |
|    | ΔE=-0.889 (-0.992) eV | ΔE=-0.773 (-0.773) eV | ΔE=-0.565 (-0.553) eV |
| Zr | $a_T$= 3.69 (3.61) Å | $a_T$= 3.81 (3.71) Å | $a_T$= 3.98 (3.86) Å |
|    | ΔE=0.543 (0.525) eV | ΔE=0.417 (0.412) eV | ΔE=0.288 (0.312) eV |
| Hf | $a_T$= 3.65 (3.56) Å | $a_T$= 3.77 (3.57) Å | $a_T$= 3.99 (3.86) Å |
|    | ΔE=0.648 (0.620) eV | ΔE=0.520 (0.505) eV | ΔE=0.378 (0.391) eV |

243 TABLE II. VBM ($E_v$), CBM ($E_c$), band gaps ($E_g$) and work functions (WF) of TMDs calculated by

244 PBE-SOC. "D" and "I" in the parenthesis after $E_g$ means "direct" and "indirect" band gaps. Due to

245 the energy level splitting at VBM caused by SOC, the calculated $E_g$ by PBE-SOC is smaller than

246 that by PBE without SOC.

| TMDs | $E_v$ | $E_c$ | $E_g$(D/I) | WF |
|---|---|---|---|---|
| MoS$_2$ | -5.86 | -4.27 | 1.59 (D) | -5.07 |
| MoSe$_2$ | -5.23 | -3.90 | 1.32 (D) | -4.57 |
| MoTe$_2$ | -4.76 | -3.83 | 0.94 (D) | -4.29 |
| WS$_2$ | -5.50 | -3.96 | 1.54 (D) | -4.73 |
| WSe$_2$ | -4.87 | -3.54 | 1.32 (D) | -4.21 |
| WTe$_2$ | -4.44 | -3.69 | 0.74 (D) | -4.06 |
| ZrS$_2$ | -6.79 | -5.71 | 1.08 (I) | -6.25 |
| ZrSe$_2$ | -6.15 | -5.86 | 0.29 (I) | -6.00 |
| ZrTe$_2$ | -4.97 | -5.69 | -0.72 (I) | -4.85 |
| HfS$_2$ | -6.83 | -5.59 | 1.23 (I) | -6.21 |
| HfSe$_2$ | -6.17 | -5.72 | 0.45 (I) | -5.94 |
| HfTe$_2$ | -4.91 | -5.53 | -0.62 (I) | -4.70 |

247

248 **Figure Captions**

249 FIG. 1. (Color online) The atomic configurations of trigonal prismatic coordination (a) and

250 octahedral coordination (b), and the corresponding *d*-orbitals splittings in trigonal prismatic and

251 octahedral crystal fields. "TM" and "X" in atomic structures represent transition metal atoms and

252 chalcogen atoms.

253

254 FIG. 2. (Color online) Band alignment of monolayer semiconducting TMDs and monolayer $SnS_2$.

255 CBM and VBM calculated by PBE-SOC are indicated by the filled grey columns, with GW

256 corrected band edges indicated by the narrower olive columns. The Fermi level is indicated by the

257 blue horizontal line and the vacuum level is at 0 eV. $SnS_2$, as a semiconducting 2D material, is

258 also listed.

259

260 FIG. 3. (Color online) Electronic structures of *H*-monolayer $MoS_2$ (a) and *T*-monolayer $ZrS_2$ (b)

261 with projected charge distribution at band edges. Yellow, violet and brown balls represent S, Mo,

262 and Zr atoms, respectively. "C" and "V" in parenthesis (for example K(C) and K(V)) mean the

263 conduction band edge and valence band edge at K point

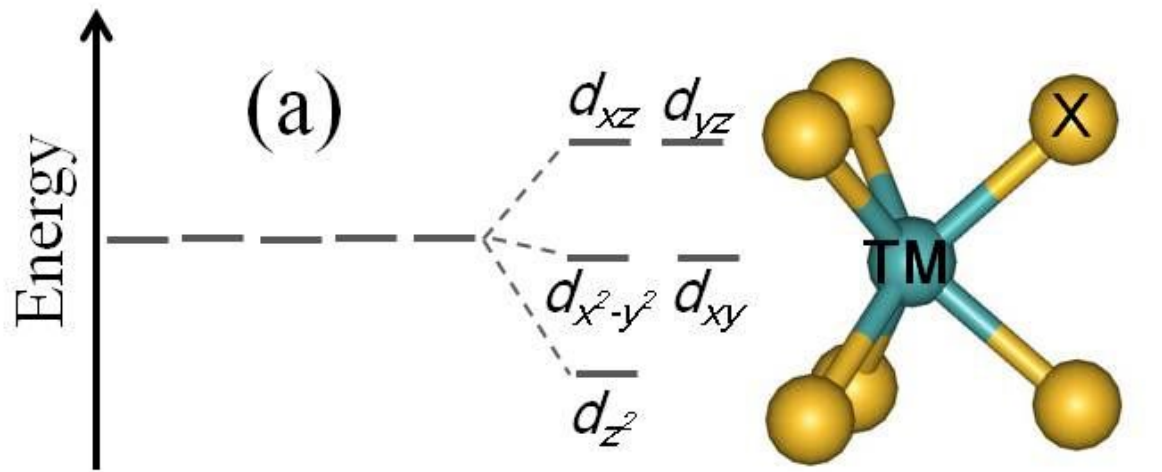
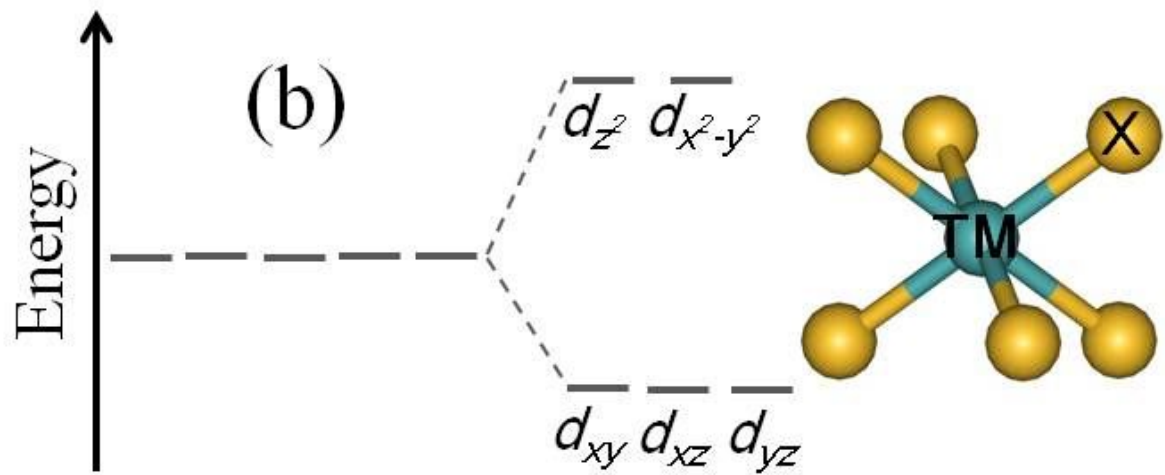

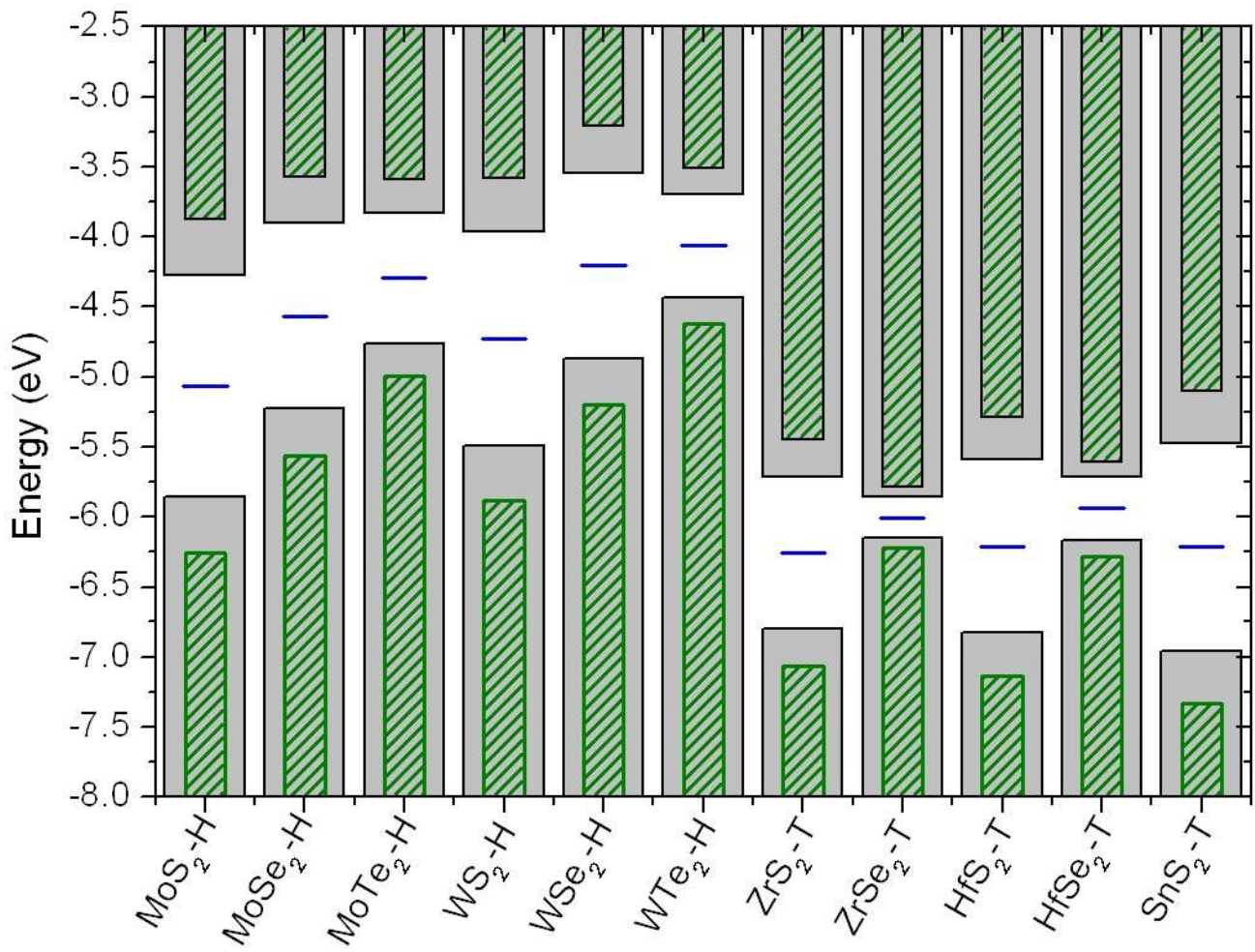

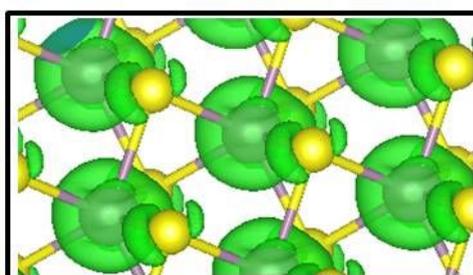

ΓK(C): 30.5%-Mo-$d z^2$; 24.8%-Mo-$d(x^2-y^2)$; 24.6%-Mo-$dxy$; 6.6%-S-$px$; 6.6%-S-$py$; 5.3%-S-$pz$.

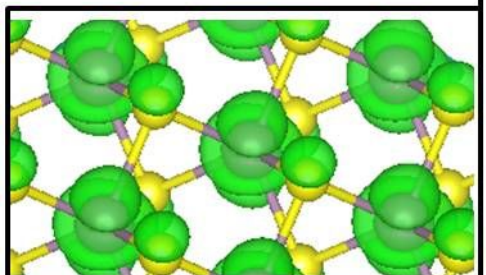

Γ(V): 78.9%-Mo-$dz^2$; 19.1%-S-$pz$.

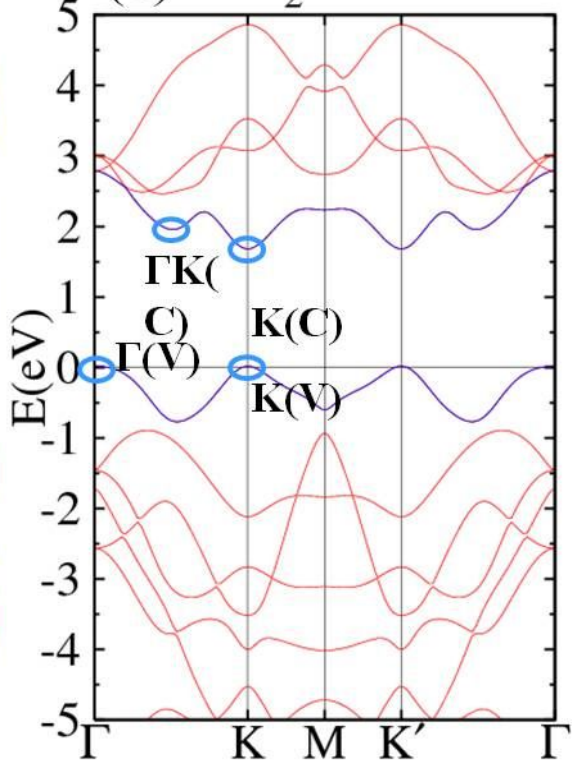

(a) MoS$_2$-*H*-PBE

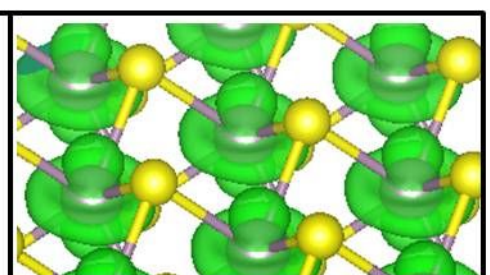

K(C): 87.7%-Mo-$dz^2$; 4.7%-Mo-$s$; 4.0%-S-$px$; 4.1%-S-$py$.

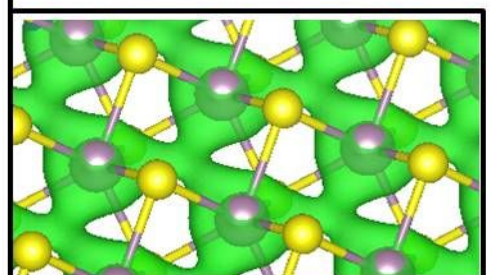

K(V): 42.1%-Mo-$d(x^2-y^2)$; 41.9%-Mo-$dxy$; 7.7%-S-$px$; 7.7%-S-$py$.

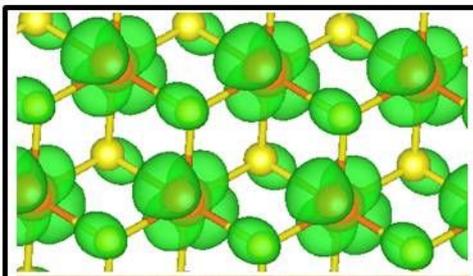

M(C): 33.9%-Zr-$dxz$; 33.4%-Zr-$dz^2$; 12.3%-Zr-$d(x^2-y^2)$; 14.3%-S-$pz$.

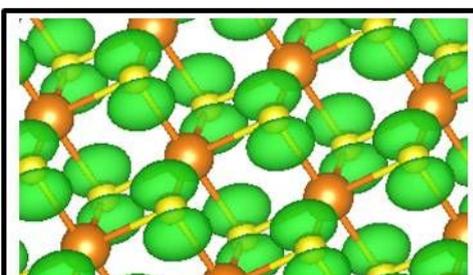

M(V): 87.3%-S-$py$; 12.7%-Zr-$py$.

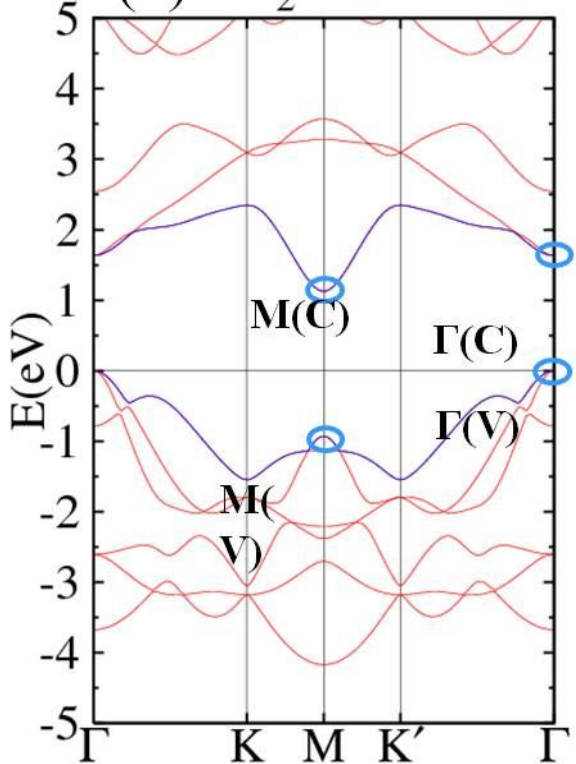

(b) ZrS$_2$-*T*-PBE

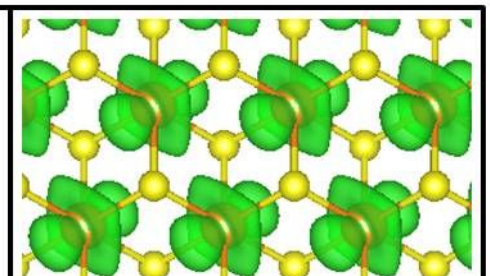

Γ(C): 66.2%-Zr-$dxz$; 32.4%-Zr-$d(x^2-y^2)$.

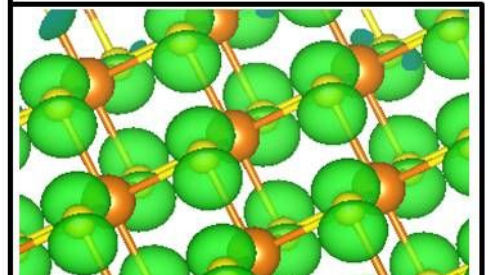

Γ(V): 95.4%-S-$py$; 4.5%-Zr-$py$.